\documentclass[prb,showpacs,preprintnumbers,amsmath,amssymb,twocolumn,superscriptaddress,longbibliography]{revtex4-1}
\usepackage{graphicx}
\def\be{\begin{equation}}
\def\ee{\end{equation}}
\def\bea{\begin{eqnarray}}
\def\eea{\end{eqnarray}}

\begin{document}

\title{Pairing symmetry of an intermediate valence superconductor CeIr$_3$ investigated using $\mu$SR measurements}

\author{D. T.  Adroja} 
\email{devashibhai.adroja@stfc.ac.uk}
\affiliation{ISIS Facility, Rutherford Appleton Laboratory, Chilton, Didcot Oxon, OX11 0QX, United Kingdom} 
\affiliation{Highly Correlated Matter Research Group, Physics Department, University of Johannesburg, PO Box 524, Auckland Park 2006, South Africa}
\author{A. Bhattacharyya}
\email{amitava.bhattacharyya@rkmvu.ac.in} 
\address{Department of Physics, Ramakrishna Mission Vivekananda Educational and Research Institute, Belur Math, Howrah 711202, West Bengal, India}
\author{Y. J. Sato}
\affiliation{Graduate School of Engineering, Tohoku University, Sendai 980-8577, Japan} 
\affiliation{Institute for Materials Research, Tohoku University, Oarai, Ibaraki 311-1313, Japan}
\author{M. R. Lees}
\affiliation{Department of Physics, University of Warwick, Coventry CV4 7AL, United Kingdom} 
\author{P. K. Biswas} 
\affiliation{ISIS Facility, Rutherford Appleton Laboratory, Chilton, Didcot Oxon, OX11 0QX, United Kingdom}  
\author{K. Panda}
\address{Department of Physics, Ramakrishna Mission Vivekananda Educational and Research Institute, Belur Math, Howrah 711202, West Bengal, India}
\author{Gavin B. G. Stenning}
\address{ISIS Facility, Rutherford Appleton Laboratory, Chilton, Didcot Oxon, OX11 0QX, United Kingdom}
\author{A. D. Hillier} 
\affiliation{ISIS Facility, Rutherford Appleton Laboratory, Chilton, Didcot Oxon, OX11 0QX, United Kingdom}
\author{D. Aoki}
\affiliation{Institute for Materials Research, Tohoku University, Oarai, Ibaraki 311-1313, Japan}

\date{\today}
\begin{abstract}

We have investigated the bulk and microscopic properties of the rhombohedral intermediate valence superconductor CeIr$_3$ by employing magnetization, heat capacity, and muon spin rotation and relaxation ($\mu$SR) measurements. The magnetic susceptibility indicates bulk superconductivity below $T_\mathrm{C} = 3.1$~K. Heat capacity data also reveal a bulk superconducting transition at $T_\mathrm{C} = 3.1$~K with a second weak anomaly near 1.6~K. At $T_{\mathrm{C}}$, the jump in heat capacity $\Delta C$/$\gamma T_{\mathrm{C}} \sim 1.39(1)$, is slightly less than the BCS weak coupling limit of 1.43. Transverse-field $\mu$SR measurements suggest a fully gapped, isotropic, $s$-wave superconductivity with 2$\Delta(0)/k_{\mathrm{B}}T_{\mathrm{C}} = 3.76(3)$, very close to 3.56, the BCS gap value for weak-coupling superconductors. From the temperature variation of magnetic penetration depth, we have also determined the London penetration depth $\lambda_{\mathrm{L}}(0) = 435(2)$~nm, the carriers' effective mass enhancement $m^{*} = 1.69(1)m_{\mathrm{e}}$ and the superconducting carrier density $n_{\mathrm{s}} = 2.5(1)\times 10^{26}$ carriers m$^{-3}$. The fact that LaIr$_3$, with no $4f$-electrons, and CeIr$_3$ with $4f^{n}$ electrons where $n \le 1$-electron (Ce ion in  a valence fluctuating state), both exhibit the same $s$-wave gap symmetry indicates that the physics of these two compounds is governed by the Ir-$d$ band near the Fermi-level, which is in agreement with previous band structure calculations.

\end{abstract}

\pacs{71.20.Be, 75.10.Lp, 76.75.+i}

\maketitle

\section{Introduction}

\noindent The strongly correlated electron systems of Ce, Yb, and U have attracted considerable attention in condensed matter physics, both theoretically and experimentally, due to the observation of heavy fermion (HF) and valence fluctuation behavior, unconventional superconductivity, quantum criticality, and spin and charge gap formation~\cite{Coleman}. The great interest in heavy fermion systems originated with the identification of superconductivity in CeCu$_2$Si$_2$ with $T_{\mathrm{C}} =0.7$~K in 1979 by Steglich {\it et al}~\cite{Steglich1979}. At that time it was thought that magnetism and superconductivity would not occur simultaneously. Nevertheless, in CeCu$_2$Si$_2$, the 4$f$ electrons which give rise to the local magnetic moments also seem to be responsible for the unconventional superconductivity~\cite{Smidmann2018}. Unconventional superconductivity was also reported in other Ce-based heavy fermion compounds including CeCoIn$_5$, which has a $T_{\mathrm{C}}$ of 2.3 K~\cite{Petrovic2001}, and the noncentrosymmetric HF superconductor CePt$_{3}$Si~\cite{Bauer2004}, a system without a center of inversion in the crystal structure that exhibits a coexistence of antiferromagnetic order ($T_{\mathrm{N}}=2.2$~K) and superconductivity ($T_{\mathrm{C}}=0.75$~K). Usually, the conventional BCS theory of superconductivity does not apply to these exotic systems~\cite{Bardeen}. Heavy fermions have a diverse range of ground states including superconductors such as UBe$_{13}$~\cite{McElfresh1990} and UPt$_3$~\cite{Visser1983,UPt3} which both adopt unconventional superconducting ground states. There are many magnetic HF systems which exhibit unconventional superconductivity under applied pressure. For example, CeIn$_{3}$ ($T_\mathrm{C} = 1.2$~K at 2.46~GPa)~\cite{Knebel2001}, CePd$_{2}$Si$_{2}$ ($T_\mathrm{C} = 0.43$~K, at 30~GPa)~\cite{Mathur1998}, CeRh$_{2}$Si$_{2}$ ($T_\mathrm{C} = 0.35$~K at 0.9~GPa)~\cite{Movshovich2002} and Ce$TX_{3}$ ($T=$~Co, Rh, Ir, $X=$~Si and Ge; $T_\mathrm{C} = 0.7-1.3$~K, 1-22 GPa)~\cite{Muro1998,Kimura2005,Kimura2007,Suqitani2006,Okuda2007, Settai2007,Knebel2009,Thamizhavel2005,Kawai2008,Honda2010}. All these HF superconductors have very high upper critical fields and some of them exhibit anisotropic behavior. Furthermore, it is reported that the superconductivity in CeIn$_{3}$ (under pressure $T_{\mathrm{C}}=0.2$~K at 2.5~GPa) and CeCoIn$_5$ has $d$-wave pairing symmetry, mainly induced by the antiferromagnetic spin fluctuations, in a way that is very similar to the high-temperature cuprates.  Strong interest in heavy fermions is also generated by the similarities seen in the phase diagrams of HF superconductors and high-temperature superconductors, including the cuprates and Fe-based materials~\cite{Grauel,Si,Zhao,Paglione}, where spin fluctuations are also suggested to play an important role.

\begin{figure*}[t]
\centering
\includegraphics[height=0.4\linewidth,width=0.9\linewidth]{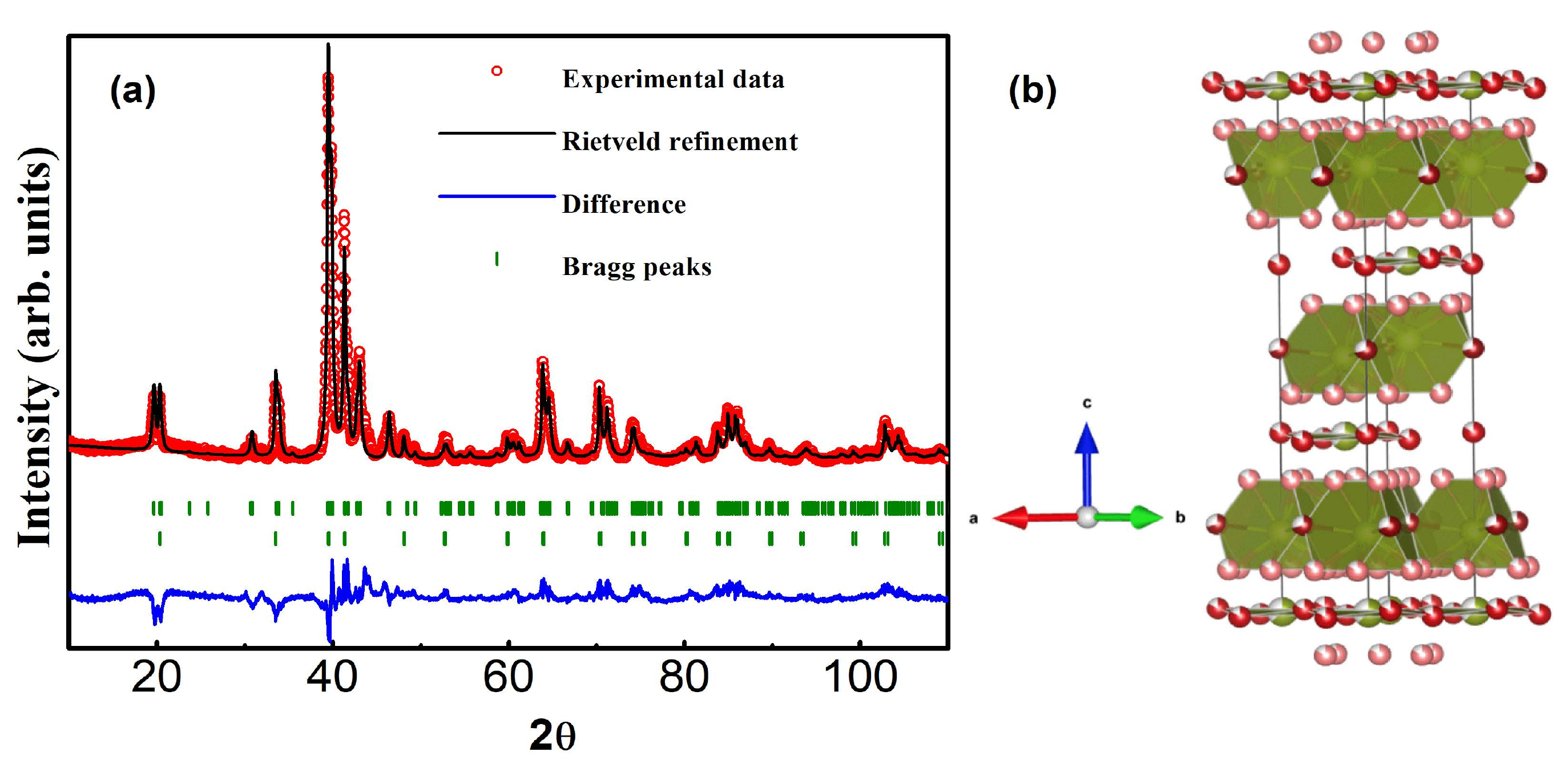}
\caption{(Color online)(a) Rietveld refinement of the X-ray diffraction pattern of CeIr$_{3}$. The data are shown as red circles, and the result of the refinement as a solid line (black). We have used rhombohedral phase (space group $R\bar{3}m$, No. 166) of CeIr$_{3}$  as a main phase and added CeIr$_{2}$ cubic  phase (space group $Fd\bar{3}m$,  No. 227) as an impurity phase. The vertical green bars show the Bragg peaks' positions, top for CeIr$_{3}$ phase and bottom for CeIr$_{2}$ phase. (b) Rhombohedral crystal structure of CeIr$_{3}$ where the Ce atoms are the bigger spheres, and the Ir atoms are the smaller spheres.}
\label{xrd}
\end{figure*}

\noindent Recently, $R$Ir$_{3}$ ($R =$~La and Ce) based materials have attracted considerable attention both experimentally and theoretically due to the observation of superconductivity with strong spin-orbit coupling~\cite{Hal, Sato2018, BhattacharyyaLaIr3}. CeIr$_3$ forms in a PuNi$_3$-type rhombohedral crystal structure (Fig.1), space group $R\bar{3}m$ (166, $D^5_{\mathrm{3d}}$ )~\cite{Sato2018}. Sato {\it et al.}~\cite{Sato2018} reported bulk type-II superconductivity in HF CeIr$_3$, with a $T_{\mathrm{C}} = 3.4$~K which is the second highest $T_{\mathrm{C}}$ among the Ce-based HF compounds. The crystal structure consists of two non-equivalent Ce sites (Ce1 and Ce2) and three Ir sites (Ir1, Ir2, and Ir3) (Fig.1b).  Gornicka {\it et al.}~\cite{Karolina} calculated the band structure of CeIr$_3$  which confirmed a non-magnetic ground state, with a small contribution from the Ce 4$f$ shell. It was reported that the density of states (DOS) at the Fermi surface principally arises from the 5$d$ states of the Ir atoms, suggesting that CeIr$_3$ is indeed an Ir 5$d$-band superconductor and that the 5$d$ electrons play a crucial role in the superconductivity.

The isostructural compound LaIr$_3$, with $T_{\mathrm{C}} = 2.5$~K, is another of the few materials~\cite{Hal, Sato2018, BhattacharyyaLaIr3} with 5$d$-electrons that exhibits superconductivity. Here as well, the bands at the Fermi surface are dominated by the Ir 5$d$ states with spin-orbit coupling, without any contribution from the La-orbitals; a similar situation is observed for CeRu$_2$~\cite{Hakimi}.

\noindent Very recently, we have investigated the superconducting properties of LaIr$_3$ using transverse-field (TF) and zero-field (ZF) muon spin rotation and relaxation ($\mu$SR) measurements. Our TF-$\mu$SR measurements revealed a fully gapped isotropic $s-$wave superconductivity with a gap to $T_{\mathrm{C}}$ ratio, 2$\Delta(0)/k_{\mathrm{B}}T_{\mathrm{C}} = 3.31$, which is smaller then the value expected from the BCS theory of 3.56, implying weak-coupling superconductivity~\cite{BhattacharyyaLaIr3}. Moreover, zero-field-$\mu$SR measurements show there are no spontaneous magnetic field below $T_{\mathrm{C}}$, which confirms time-reversal symmetry is preserved in LaIr$_3$~\cite{BhattacharyyaLaIr3}.

\begin{figure*}[t]
\centering
\includegraphics[height=0.6\linewidth,width=0.9\linewidth]{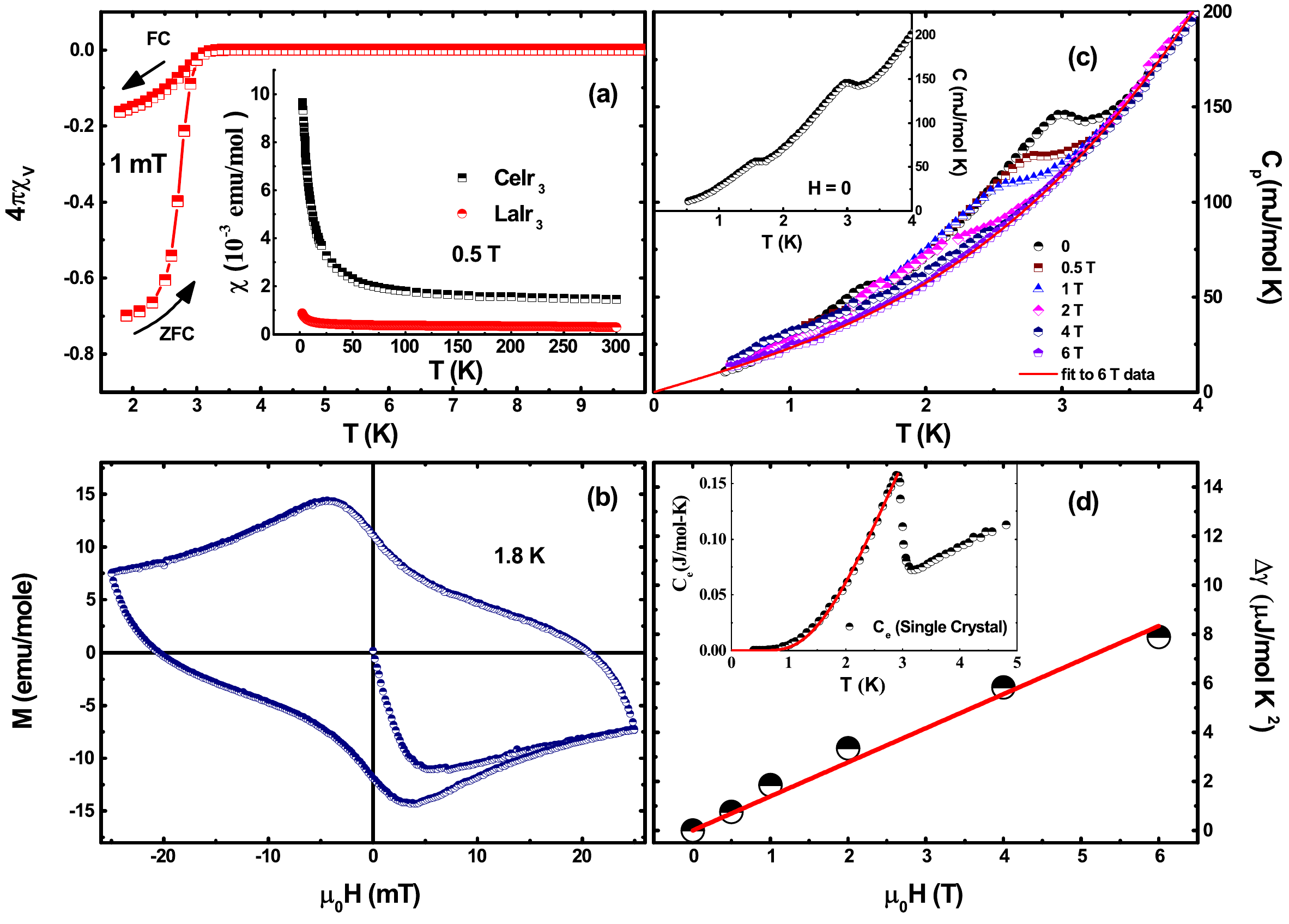}
\caption{(Color online)(a) DC magnetic susceptibility of CeIr$_3$ as a function of temperature in zero-field cooled (ZFC) and field-cooled cooling (FCC) mode. The inset shows the high-field magnetic susceptibility of CeIr$_3$ (black squares) and LaIr$_3$ (red circles). (b) Isothermal magnetic field dependence of the magnetization of CeIr$_3$ at 1.8~K. (c) Temperature dependence of the heat capacity of CeIr$_3$ at different applied fields. The solid line shows a fit to the 6~T data. The inset shows the temperature variation of the heat capacity in zero applied magnetic field. (d) Magnetic field dependence of the electronic specific heat $\Delta\gamma=[\gamma(H)-\gamma(0)]$ of CeIr$_3$ polycrystal sample extrapolated to $T\sim 0$~K. The solid line shows a linear behaviour indicating isotropic s-wave gap symmetry.  Inset of (d) shows the electronic heat capacity for CeIr$_3$ single crystal from Ref. ~\cite{Sato2018} presented here for the comparison with the polycrystal data. The solid red line represents fully gapped superconductivity.}
\label{mtrtcp}
\end{figure*}

\noindent Here we have investigated the superconducting states of the mixed valence metal CeIr$_3$ by means of magnetization, heat capacity, and TF/ZF-$\mu$SR measurements. The temperature dependence of the magnetic penetration depth, determined by TF-$\mu$SR measurements, implies a fully gapped isotropic $s-$wave nature for the superconducting state. The ZF-$\mu$SR data show no evidence of any spontaneous internal fields developing at and below $T_{\mathrm{C}}$ in the superconducting state, suggesting that time-reversal symmetry is preserved in the superconducting state of CeIr$_3$.

\section{Experimental Details}

\noindent A polycrystalline sample of CeIr$_3$ was prepared in a tetra arc furnace by arc melting stoichiometric quantities of the starting elements (Ce: 99.9 wt\%; Ir: 99.999 wt\%). The ingot was flipped and remelted five times, and the sample was quenched. The sample was subsequently annealed at 900$^{0}$ for 6 days under a vacuum of $1 \times 10^{-4}$ Pa in a quartz ampoule.  The sample was wrapped in tantalum (Ta) foil during the annealing. The sample was heated to 900$^{0}$C and held at this temperature for 6 days and then quenched by switching off the furnace. The quality of the sample was verified through powder X-ray diffraction using a Panalytical X-Pert Pro diffractometer. The temperature and field dependence of magnetization was measured using a Quantum Design Magnetic Property Measurement System SQUID magnetometer; heat capacity down to 500~mK was measured using a Quantum Design Physical Property Measurement System with a $^3$He insert. To examine the superconducting pairing symmetry and microscopic superconducting properties of CeIr$_3$, we performed TF/ZF $\mu$SR experiments at the muon beam line of the ISIS Pulsed Neutron and Muon Facility at the Rutherford Appleton Laboratory, United Kingdom using the MUSR spectrometer~\cite{Lee1999}. The powder sample of CeIr$_3$ was mounted on a silver plate (99.995\%) using GE varnish diluted with ethanol and covered with a silver foil. The sample was cooled to 50~mK using a dilution refrigerator. 100\% spin-polarized positive muons were implanted into the sample and the asymmetry of the resulting decay positrons was estimated using, $P_{\mathrm{z}}\left(t\right) = [{N_{\mathrm{F}}\left(t\right)-cN_{\mathrm{B}}\left(t\right)}]/[{N_{\mathrm{F}}\left(t\right)+cN_{\mathrm{B}}\left(t\right)}]$, where $N_{\mathrm{B}}\left(t\right)$ and $N_{\mathrm{F}}\left(t\right)$ are the number of positrons counted in the backward and forward detectors respectively and $c$ is a instrumental calibration constant determined in the normal state with a small (2~mT) transverse magnetic field. The TF-$\mu$SR data were collected in the different temperatures between 0.05 and 4~K in the presence of a 40~mT ($> \mu_{0}H_\mathrm{c1}(0) = 5.1(2)$~mT) magnetic field. ZF data were collected between 0.05 and 4~K. To reduce the impact of the magnetic fields at the sample position, correction coils were used which assured the stray fields were always less than 1~$\mu$T. All the $\mu$SR data were analyzed using WiMDA, a muon data analysis program~\cite{Pratt2000}.
 
\begin{figure*}[t]
\centering
\includegraphics[height=0.4\linewidth,width=\linewidth]{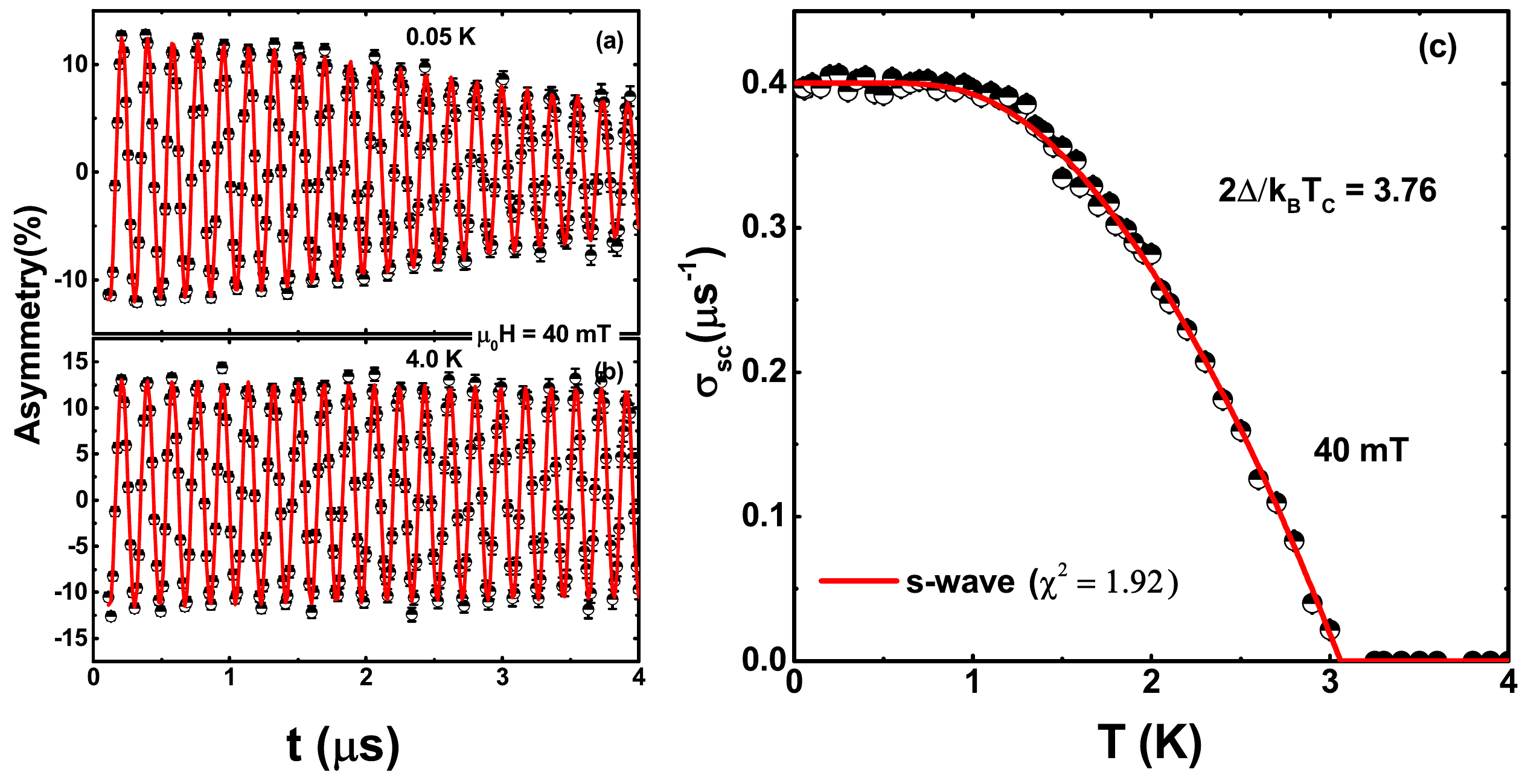}
\caption{(Color online) TF-$\mu$SR spin precession signals for CeIr$_3$ collected in an applied transverse-magnetic field of $\mu_\mathrm{0}H = 40$~mT. Asymmetry versus time in (a) the superconducting state at 0.05~K and (b) the normal state at 4.0~K. Solid lines represent fits to the data using Eq.~\ref{MuonFit1}. (c) Temperature variation of the Gaussian superconducting relaxation rate $\sigma_{\mathrm{sc}}(T)$. The line is a fit to the data using an isotropic, fully gapped $s$-wave model using Eq.~\ref{MuonFit2}.}
\label{gap}
\end{figure*}

\section{Results and Discussion}
\subsection{Crystal Structure and Physical properties}

\noindent Fig.\ref{xrd}~(a) presents the X-ray diffraction (XRD) pattern (symbols) and the Rietveld refinement fit (solid line) of the studied CeIr$_{3}$ compound. CeIr$_{3}$  crystallizes in the PuNi$_{3}$-type rhombohedral structure with the space group $R\bar{3}m$. Our analysis of the XRD data reveals an impurity phase of CeIr$_2$ with a cubic strcture (space group $Fd\bar{3}m$,  No. 227) (see Fig.1a).   The schematic unit cell obtained from Rietveld analysis of XRD data of CeIr$_{3}$ is shown in the inset of fig.\ref{xrd}(b). The lattice parameters of the synthesized CeIr$_3$ sample are $a$ =  5.2943(2) \AA~and $c$ = 26.2134(1) \AA, $\alpha = \beta = 90 \deg$ and $\gamma = 120 \deg$ which are in agreement with a previous report ~\cite{Sato2018}.  

Fig.~\ref{mtrtcp}(a) presents the temperature dependence of the magnetic susceptibility $\chi\left(T\right)$ in zero-field-cooled (ZFC) state and field-cooled (FC) state which confirms the bulk type-II superconductivity at 3.1~K in CeIr$_3$. The isothermal magnetic field dependence of magnetization at 0.4~K is shown in Fig.~\ref{mtrtcp}(b).  Fig.~\ref{mtrtcp}(c) shows the temperature dependence of the heat capacity $C_{\mathrm{P}}$ at different applied magnetic fields. The inset in Fig.~\ref{mtrtcp}(c) shows the temperature variation of heat capacity at zero applied magnetic field. A clear signature of a superconducting transition is observed below 3.1~K in $C_{\mathrm{P}}\left(T\right)$ data. Furthermore, another weak transition in $C_{\mathrm{P}}\left(T\right)$ is seen below 1.6~K. As single crystal heat capacity of CeIr$_{3}$ shows only one transition at $T_\mathrm{C} = 3.1$~K as shown in inset of Fig.~\ref{mtrtcp}(d), and no sign of second transition~\cite{Sato2018}, which might suggest that the second transition observed in the polycrystalline CeIr$_{3}$ near 1.6~K might be related to a second superconducting impurity phase or very small variation on Ir composition (i.e. inhomogeneous Ir composition, CeIr$_{3-\delta}$)~\cite{Hal}. Further, to check whether the transition near 1.6~K  is coming from CeIr$_{2}$ phase or not, we have synthesised CeIr$_{2}$ polycrystalline sample and carried out XRD and heat capacity $C_{\mathrm{P}}$ (down to 400~mK) measurements. The results of XRD and $C_{\mathrm{P}}$ of CeIr$_{2}$ sample are given in the supplimentary material (SM).~\cite{SM}  The $C_{\mathrm{P}}$ of CeIr$_{2}$ decreases with decreasing temperature from 2.5~K dwon to 400~mK and does not reveal any anomaly/peak due to an onset of superconductivity. This confirms that the weak anomaly observed in the heat capacity data of CeIr$_3$ near 1.6~K is not coming from the  CeIr$_{2}$ phase and need further investigation. The jump in the heat capacity of CeIr$_3$ is suppressed in a magnetic field of 6~T. The heat capacity data were fitted using $C_{\mathrm{P}}\left(T\right)/T = \gamma +\beta T^2$ where $\gamma$ and $\beta$ are electronic specific heat coefficient and lattice specific heat coefficient, respectively.  The least squares fit yields $\gamma = 21.66(2)$~mJ/(mol-K$^2$), $\beta = 1.812(1)$~mJ/(mol-K$^4$), and then using $\beta = {nN_A}\frac{12}{5}\pi^4R\Theta_D^{-3}$, where $R = 8.314$~J/mol-K is the universal gas constant, $n$ is the number of atoms per formula unit, and $N_\mathrm{A}$ is Avogadro's number, we estimate that the Debye temperature $\Theta_D = 162(2)$~K. Sato {\it et al.} have reported  the heat capacity jump $\Delta C_{\mathrm{P}}/\gamma T_{\mathrm{C}} \sim 1.39(1)$ and 2$\Delta(0)$/$k_{\mathrm{B}}T_{\mathrm{C}}$ = 3.83(1)~\cite{Sato2018} for a single crysral of CeIr$_3$, which is closer to the theoretical BCS limit of a weak-coupling superconductor (3.56). Both of these values suggest that CeIr$_3$ can be categorized as a weak-coupling superconductor. Fig.~\ref{mtrtcp}(d) presents the magnetic field dependence of the electronic specific heat $\Delta \gamma[=\gamma(H)-\gamma(0)]$ extrapolated to $T\sim 0$~K. The linear trend follows the behavior expected for a conventional BCS type superconductor. From the exponential dependence of $C_e$ of CeIr$_{3}$ single crystal, as shown in the inset of Fig.~\ref{mtrtcp}(d), we obtained 2$\Delta(0)$/$k_{\mathrm{B}}T_{\mathrm{C}}$ to be 3.81. The inset in Fig.~\ref{mtrtcp}(a) shows the high-field magnetic susceptibility measured up to 300~K for both CeIr$_3$ and LaIr$_3$. The susceptibility of CeIr$_3$ is higher than that of LaIr$_3$ and exhibits considerable temperature dependence below 25~K. The high temperature (50-300~K), weak temperature dependence, behavior of the susceptibility of CeIr$_3$ indicates the presence of strong hybridization between localized 4f-electron and conduction electrons and the mixed valence of the Ce ions. The low-temperature rise could be attributed to the Curie tail from an impurity. An X-ray photoelectron spectroscopy study reported that the Ce ions have a strongly intermediate valence character in CeIr$_3$~\cite{Karolina}. The Ce ion valence of 3.6 in CeIr$_3$ was estimated using the superconducting transition temperatures, $T_{\mathrm{C}}$, of the pseudo-binaries of the isostructural compounds LaIr$_3$, CeIr$_3$, and ThIr$_3$~\cite{Hakimi}. Furthermore, evidence of an intermediate valence, between $3^+$ and $4^+$, of the Ce ions in CeIr$_3$ comes from Vegard's law by plotting the volume versus covalent radius of the $R^{3+}$ metal in the $R$Ir$_3$ series. The volume increases monotonically with an increase in the radius, except for Ce, for which the unit cell volume is much smaller and comparable with the unit cell volume of GdIr$_3$ supporting the intermediate valence of Ce ion in CeIr$_3$~\cite{Karolina}. 

\subsection{Superconducting gap structures}

The TF-$\mu$SR asymmetry spectra measured in an applied magnetic field of 40~mT are displayed in Figs.~\ref{gap}(a-b). The data in Fig.~\ref{gap}(a) were taken at the base temperature in the superconducting state and in Fig.~\ref{gap}(b) at a higher temperature, well into the normal state. At $T\ge T_{\mathrm{C}}$, the muon asymmetry oscillates with a very small damping, suggesting that the internal field distribution is extremely uniform. On the other hand, the asymmetry spectrum measured at $T\le T_{\mathrm{C}}$ shows an increased in damping, suggesting an inhomogeneous field distribution due to the vortex state. To obtain quantitative information about the superconducting state in CeIr$_3$, we first tried to analyze TF-$\mu$SR data recorded at various temperatures using two Gaussian components, one to account CeIr$_3$ phase and another to account the impurity phase. However, two componets model gave unphysical values of the parameters and fit did not converge. We therefore fitted our TF-$\mu$SR data using a single Gaussian model~\cite{Bhattacharyya1, Bhattacharyya2, Adroja, Bhattacharyyarev} given by,
\begin{equation}
G_{x}(t) = C_{1}\cos(\omega_{1}t+\Phi)\exp\bigg(\frac{-\sigma^{2}t^{2}}{2}\bigg)+C_{2}\cos(\omega_{2}t+\Phi),
\label{MuonFit1}
\end{equation}
\noindent where $C_{i}$ and $\omega_{i}$ ($i = 1$, 2), are the transverse-field asymmetries and the muon spin precision frequencies that arise from the sample and the silver sample holder (this could also include the impurity phase), and $\Phi$ and $\sigma$ are a phase factor and total Gaussian depolarization rate, respectively. During the fitting $C_\mathrm{2}$ was fixed at 35\%, its low-temperature value, and the asymmetry spectra were then fit by varying the value of $C_\mathrm{1}$ which is nearly independent of temperature. The phase, $\Phi$, was also fixed to the value obtained at low temperature. Figs.~\ref{gap}(a-b) also include fits to the data (the solid red lines) using Eq.~\ref{MuonFit1}, show a good correspondence between the experimental and the calculated asymmetry spectra.

The values of $\sigma$ determined from the fits consists of two parts; one part comes from the superconducting signal, $\sigma_{sc}$, and the other part is the nuclear magnetic dipolar contribution, $\sigma_{\mathrm{nm}}$, which is taken to be constant over the entire temperature range studied.
The superconducting depolarization rate $\sigma_{\mathrm{sc}}$ is then calculated using $\sigma_{\mathrm{sc}} = \sqrt{\sigma^{2}-\sigma_{\mathrm{nm}}^2}$. The temperature variation of $\sigma_{\mathrm{sc}}$ shown in Eq.~\ref{MuonFit1} is modeled using a standard expression within the local London approximation~\cite{ThCoC2, Bhattacharyya1, Adroja} with 
\begin{eqnarray}
\label{MuonFit2}
\frac{\sigma_{\mathrm{sc}}\left(T\right)}{\sigma_{\mathrm{sc}}\left(0\right)} &=& \frac{\lambda^{-2}\left(T,\Delta_{0}\right)}{\lambda^{-2}\left(0,\Delta_{0}\right)} \nonumber \\
 &=& 1 +  \frac{1}{\pi}\int_{0}^{2\pi}\int_{\Delta\left(T\right)}^{\infty}\left(\frac{\delta f}{\delta E}\right) \times \frac{E\mathrm{d}E\mathrm{d}\phi}{\sqrt{E^{2}-\Delta^2\left(T,\Delta_{}\right)}} \nonumber
\end{eqnarray}

\noindent where $f = [1+\exp(-E/k_{B}T)]^{-1}$ is the Fermi function, $\phi$ is the azimuthal angle in the direction of Fermi surface, and $\Delta_\mathrm{}(T,0) = \Delta_\mathrm{0}\delta(T/T_\mathrm{C})$g$(\phi)$. $\Delta_\mathrm{0}$, the gap value at zero temperature, is the only adjustable parameter. The temperature dependence of the gap can be approximated by $\delta\left(T/T_{\mathrm{C}}\right) = \tanh\left[1.82\left[1.018\left(T_{\mathrm{C}}/T-1\right)\right]^{0.51}\right]$, and g$\left(\phi\right)$ gives the angular dependence of the gap function where $\phi$ is the polar angle for the anisotropy. The spatial dependence g$(\phi$) is substituted by (a) 1 for an $s$-wave gap, and (b) $\vert\cos(2\phi)\vert$ for a $d$-wave gap with line nodes~\cite{Annett,Pang}. 

A conventional isotropically gapped model describes the data very well, as shown by the solid red line in Fig.3(c). Using this isotropic model, the refined critical temperature is $T_{\mathrm{C}} = 3.1$~K and the gap to $T_{\mathrm{C}}$ ratio of 2$\Delta(0)/k_\mathrm{B}T_{\mathrm{C}} = 3.76(3)$, is close to the value of 3.56 expected from a weak-coupling BCS theory. This value is in agreement with the heat capacity data. 

\begin{figure}[t]
\centering
\includegraphics[width=\linewidth]{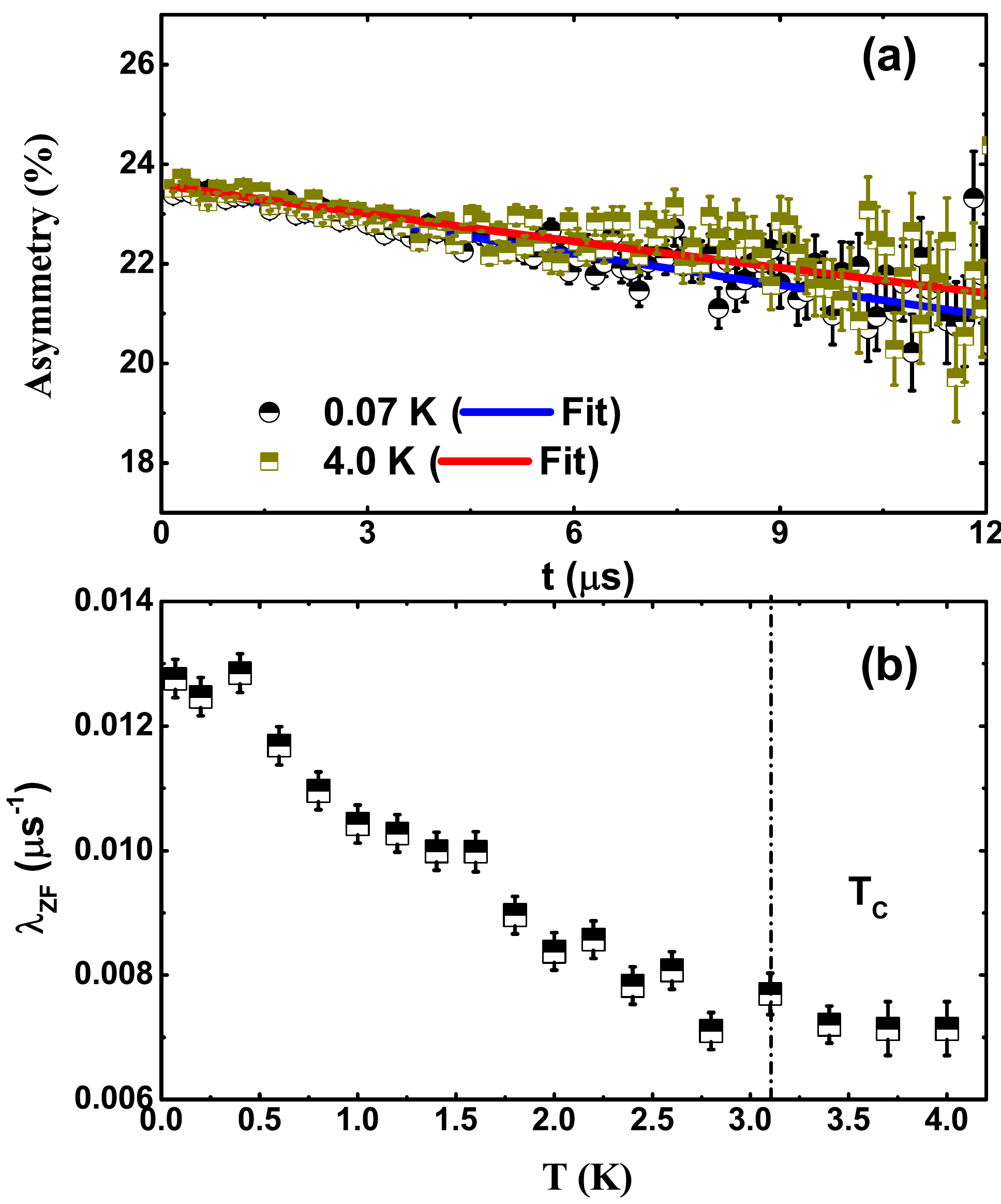}
\caption{(Color online) (a) Zero-field $\mu$SR asymmetry spectra for CeIr$_3$ collected at 0.07~K (black circles) and 4.0~K (dark yellow squares) together with lines that are least-squares fits the data using Eq.~\ref{MuonFit3}(b) Temperature variation of zero-field muon relaxation rate.}
\label{ZFasymmetry}
\end{figure}

Using the TF-$\mu$SR results, the other superconducting parameters characterizing the superconducting  ground state of CeIr$_3$ can be evaluated. For a triangular lattice~$\sigma_{sc}^{2} = \frac{0.00371 \times \phi_{0}^2}{\lambda^{4}}$ where $\phi_{0}$ is the flux quantum number 2.07$\times$ 10$^{-15}$ T m$^{2}$ and $\gamma_{\mu}$ is the muon gyromagnetic ratio, $\gamma_{\mu}/2\pi$ = 135.5 MHz T$^{-1}$. Using this relation we have estimated the magnetic penetration depth, $\lambda(0)$ = 435(2)~nm.  The  London theory~\cite{Sonier} gives  the relation between microscopic quantities, $\lambda$ (or $\lambda_{\mathrm{L}}$), effective mass ($m^{*}$) and the superconducting carrier density ($n_\mathrm{s}$ );  $\lambda_{\mathrm{L}}^2$=$\lambda^2$ = $\frac{m^{*}c^{2}}{4\pi n_\mathrm{s}e^{2}}$, here $m^{*} = \left(1+\lambda_\mathrm{e-ph}\right)m_\mathrm{e}$, where   $\lambda_{\mathrm{e-ph}}$ is  the electron-phonon coupling constant and ${m_\mathrm{e}}$ is an electron mass . Using McMillan's relation~\cite{McMillan}, $\lambda_{\mathrm{e-ph}}$ can be determined using  

\begin{equation}
\label{McMillanEq}
\lambda_\mathrm{e-ph} = \frac{1.04+\mu^{*}\ln(\Theta_\mathrm{D}/1.45T_\mathrm{C})}{(1-0.62\mu^{*})\ln(\Theta_\mathrm{D}/1.45T_\mathrm{C})-1.04},
\end{equation}

\noindent where $\Theta_{\mathrm{D}}$ is the Debye temperature. Assuming a repulsive screened Coulomb parameter $\mu^{*} = 0.13$~\cite{Allen}, we have estimated $\lambda_{\mathrm{e-ph}} = 0.57(2)$. This value of $\lambda_{\mathrm{e-ph}}$ is larger than 0.02 to 0.2 observed for many Fe-based superconductors (11- and 122-family) and cuprates (YBCO-123) ~\cite{AMZhang2013}, but smaller than 1.38 for LiFeAs ~\cite{LaFeAs},  1.53 for PrFeAsO$_{0.60}$F$_{0.12}$~\cite{PrFeAsOF} and 1.2 for LaO$_{0.9}$F$_{0.1}$FeAs ~\cite{LaOFFeAs}. Given CeIr$_3$ is a type II superconductor,  and using the above estimated value of  $\lambda_{\mathrm{e-ph}}$ and $\lambda_{\mathrm{L}}$, we have estimated the effective-mass enhancement $m^{*} = 1.69(1) m_{\mathrm{e}}$ and superconducting carrier density $n_{\mathrm{s}} = 2.5(1) \times 10^{26}$ carriers $m^{-3}$. The superconducting parameters of CeIr$_3$ and LaIr$_3$ are collected together in Table~\ref{Comparison}. 

\begin{table}
\caption{Superconducting parameters of CeIr$_{3}$ and LaIr$_{3}$. The parameters values of LaIr$_{3}$ comes from Ref.~\cite{BhattacharyyaLaIr3}  }
\label{Comparison}
\centering
\begin{tabular}{lcc}
\hline
Parameter & CeIr$_3$ &  LaIr$_3$ \\  \hline
$T_\mathrm{C}$ (K)     & 3.1 & 2.5 \\
$\mu_{0}H_\mathrm{c1}$(mT)  & 5.1(2)& 11.0(2)\\
$\mu_{0}H_\mathrm{c2}$(T) &  4.65(3) &  3.84(2)\\
$\gamma$(0) (mJ/mol K$^{2}$) &  21.66(2) &15.32(3)  \\
$\Theta_\mathrm{D}$ & 162(2) & 430(4)  \\
$\Delta C/\gamma T_\mathrm{C}$ & 1.39(1) &  1.0(2) \\ 
2$\Delta$/$k_\mathrm{B} T_\mathrm{C}$ & 3.76(3) &  3.31(1) \\
$\lambda$(nm) & 435(2) & 386(3) \\
$\lambda_\mathrm{e-ph}$ & 0.57(2) & 0.53(3) \\
$n_\mathrm{s}$(carriers/m$^{3})$ & 2.5(1) $\times$ 10$^{26}$ & 2.9(1) $\times$ 10$^{27}$\\ 
\hline
\end{tabular}
\label{tab:1}
\end{table}

\subsection{Zero-field muon spin relaxation}

ZF-$\mu$SR  muon asymmetry spectra above (dark yellow) and below (black) $T_{\mathrm{C}}$, that are representative of the data collected are shown in Figs.~\ref{ZFasymmetry}(a-b). Both spectra exhibit a slow and almost indistinguishable exponential relaxation. Fits to the ZF-$\mu$SR spectra at several temperatures between 0.07 and 4.0~K were made using the Lorentzian function~\cite{Adroja1, Adroja2, Adroja3, Bhattacharyya3},

\begin{equation}
\label{MuonFit3}
G_{\mathrm{z}}(t) =C_{\mathrm{0}} \exp ({-\lambda_{ZF} t})+C_{\mathrm{bg}},
\end{equation}

\noindent where $C_{\mathrm{0}}$, $C_{\mathrm{bg}}$ and $\lambda_{ZF}$ are the total initial asymmetry from muons probing the sample, the asymmetry arising from muons landing in the silver sample holder, and the electronic relaxation rate, respectively. The parameters $C_{\mathrm{0}}$, and $C_{\mathrm{bg}}$ are found to be temperature independent. The zero-field-$\mu$SR measurements reveal the relaxation rate between 0.07 and 4~K is slightly temperature dependent, suggesting the presence of weak spin-fluctuations. This effect is not seen in LaIr$_3$, which suggests that the spin fluctuations originate from the Ce moments that are in intermediate valence state. There is no clear change in $\lambda_{ZF}$ as the samples cools though $T_{\mathrm{C}}$ indicating that time-reversal symmetry is likely preserved in CeIr$_{3}$.\\

\section{Summary}

In summary, we have examined the superconducting properties, including the superconducting ground state, of CeIr$_3$. Magnetic susceptibility measurements show CeIr$_{3}$ is a bulk type-II superconductor with $T_{\mathrm{C}} = 3.1$~K. The heat capacity of polycrystalline CeIr$_{3}$ shows the superconducting transition near 3.1~K and a second weaker anomaly near 1.6~K. Given that the heat capacity of CeIr$_{3}$ single crystal exhibits only one transition near $T_\mathrm{C} = 3.1$~K ~\cite{Hal} and no peak observed in the heat capacity of CeIr$_{2}$ between 2.5~K and 400~mK~\cite{SM}, the second transition near 1.6~K could be associated with some variation in Ir content throughout the sample and need further investigation. The temperature dependence of the ZF-$\mu$SR relaxation rate confirmed the preservation of time-reversal symmetry below $T_\mathrm{C}$; the very weak temperature dependence suggests the presence of weak spin fluctuations. Transverse-field $\mu$SR measurements reveal the CeIr$_3$ exhibits an isotropic fully gapped $s$-wave type superconductivity with a gap to $T_{\mathrm{C}}$ ratio, 2$\Delta(0)/k_{\mathrm{B}}T_{\mathrm{C}} = 3.76$, compared to the expected BCS value of 3.56 suggesting weak-coupling superconductivity.  The $s$-wave pairing symmetry observed in both LaIr$_3$~\cite{BhattacharyyaLaIr3}, a material with no 4$f$-electrons and CeIr$_{3}$, with  less than one 4$f$-electron, indicates that the superconductivity is controlled by the Ir-$d$ bands near the Fermi level in both the compounds.   

\section{Acknowledgments}

A. B. would like to thank DST India, for an Inspire Faculty Research Grant (DST/INSPIRE/04/2015/000169). D. T. A. and A. D. H. would like to acknowledge the CMPC-STFC, grant number CMPC-09108. D. T. A. is grateful to the JSPS for an invitation fellowship. K. P. would like to acknowledge DST India, for an Inspire Fellowship (IF170620).

\clearpage
{\bf Supplementary Materials: Pairing symmetry of an intermediate valence superconductor CeIr$_3$ investigated using $\mu$SR measurements}

\section*{}
{\bf x-ray diffraction and Heat Capacity measurements of CeIr$_2$}
\\*

A polycrystalline sample of CeIr$_2$ was prepared in a tetra arc furnace by arc melting stoichiometric quantities of the starting elements (Ce: 99.9 wt\%; Ir: 99.999 wt\%). The ingot was flipped and remelted five times to improve the homogeneity. The quality of the sample was verified through powder X-ray diffraction using a Panalytical X-Pert Pro diffractometer. The x-day powder diffraction (XRD) pattern of CeIr$_2$ is shown in Fig.1 (top) along with the simulated XRD pattern (bottom) using the cubic  phase (space group $Fd\bar{3}m$,  No. 227, with cubic  lattice parameter a=5.394~{\AA}). A very good agreement between the experimental XRD pattern and the simulated pattern indicates the single phase nature of CeIr$_2$ sample.

The heat capacity of CeIr$_2$ down to 400~mK was measured using a Quantum Design Physical Property Measurement System with a $^3$He insert. Fig.2 shows the temperature dependence heat capacity of CeIr$_2$ between 2.5 K and  400~mK in zero applied field. It is clear that the heat capacity decreases with decreasing temperature down to 400~mK. No clear sign of an onset of superconductivity was observed in the heat capacity of CeIr$_2$ down to 400m~K, which suggests that superconducting transition temperature of our CeIr$_2$ sample is below 400~mK.

\begin{figure*}
\centering
\includegraphics[height=0.4\linewidth,width=0.9\linewidth]{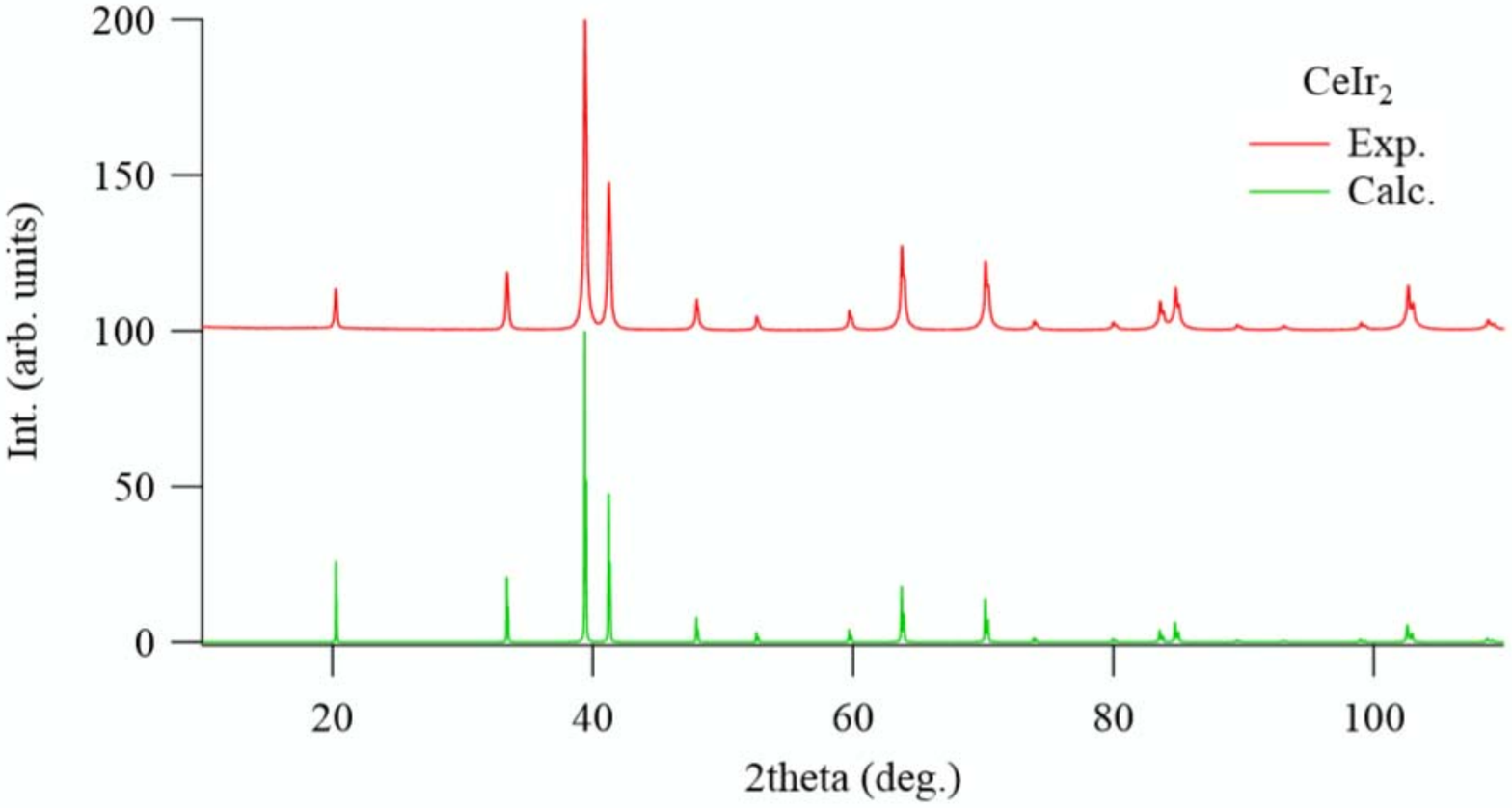}
\caption{(Color online)(top) The experimental X-ray diffraction pattern of CeIr$_2$ and (bottom) the simulated X-ray diffraction pattern of CeIr$_{2}$ using cubic  phase (space group $Fd\bar{3}m$,  No. 227).}
\label{xrd}
\end{figure*}

\begin{figure*}
\centering
\includegraphics[height=0.6\linewidth,width=0.9\linewidth]{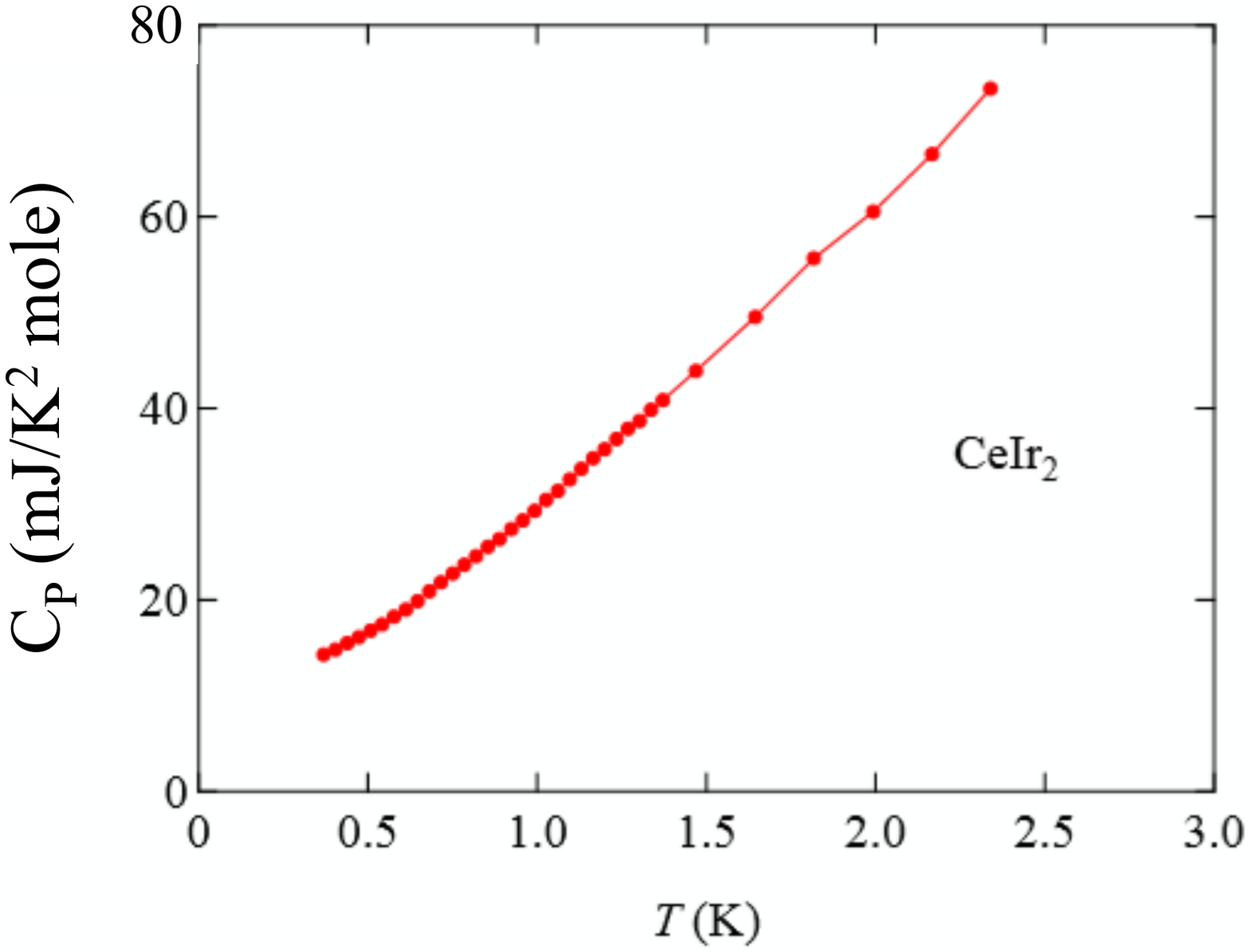}
\caption{(Color online)Temperature dependence of the heat capacity of CeIr$_2$ in zero applied field.}
\label{mtrtcp}
\end{figure*}
\end{document}